\documentclass[12pt]{iopart}

\usepackage{color}

\usepackage[english]{babel}
\usepackage{graphicx}
\usepackage{cite}
\usepackage{subfigure}

\newcommand{\D}{{\rm d}}

\begin{document}
\title{Diffusion with Optimal Resetting}

\author{Martin R. Evans$^{(1,3)}$ and Satya N. Majumdar$^{(2,3)}$}
\address{$^{(1)}$ SUPA, School of Physics and Astronomy, University of Edinburgh, Mayfield Road, Edinburgh EH9 3JZ, United Kingdom\\
$^{(2)}$ Univ. Paris-Sud, CNRS, LPTMS, UMR 8626, Orsay F-01405, France\\
$^{(3)}$ Weizmann Institute of Science}
\eads{\mailto{m.evans@ed.ac.uk,
  	satya.majumdar@u-psud.fr}
} 
\date{\today, last edited by MRE}

\begin{abstract}
We consider the mean time to absorption 
by an
absorbing target of a diffusive particle  with the addition of a process whereby the particle
is reset to its initial position with rate $r$.  We consider several
generalisations of the model of 
M.~R.~Evans and S.~N.~Majumdar (2011),
Diffusion with stochastic resetting,
Phys. Rev. Lett.  106, 160601: (i) a space dependent
resetting rate $r(x)$ ii) resetting to a  random position $z$
drawn from a resetting distribution ${\cal P}(z)$ iii) a spatial
distribution for the absorbing target $P_T(x)$.  As an example of (i)
we show that the introduction of a non-resetting window around the
initial position can reduce the mean time to absorption provided that
the intial position is sufficiently far from the target.  We address
the problem of optimal resetting, that is, minimising the mean time to
absorption for a given target distribution.  For an exponentially
decaying target distribution centred at the origin we show that a
transition in the optimal resetting distribution occurs as the target
distribution narrows.
\end{abstract}

\section{Introduction}

Search problems occur in a variety of contexts: from animal foraging
\cite{Bell} to the target search of proteins on DNA molecules
\cite{Berg81,CBVM04,BKSV09}; from internet search algorithms to the
more mundane matter of locating one's mislaid possessions.
Often search strategies involve a mixture of local steps and
long-range moves \cite{BCMSV05,LKMK08,BC09,GLWB09,RRBWOL10}.  For
human searchers at least, a natural tendency is to return to the
starting point of the search after the length of time spent searching
becomes excessive.

In a recent paper \cite{EM11} we modelled
such a strategy as a diffusion process with an
additional rate of resetting to the starting point $x_0$
with rate $r$.
Considering the object of the search to be an absorbing target at the origin,
the duration of the search  becomes the time for the  diffusing particle
to reach the origin.
Statistics such as the mean time to absorption of the process then
give a measure of the efficiency of the search strategy, defined by
the resetting rate $r$. Moreover, the model provides a system where
the statistics of absorption times can be computed exactly.

A related model, 
where searchers have some probabilistic lifetime after which
another searcher will be sent out,
has been studied by Gelenbe \cite{Gelenbe10}
and mean times to absorption  computed.
Also, in the mathematical literature the mean first
passage time for random walkers  that have  the option of restarting at the
initial position has been considered \cite{JP10}.

In \cite{EM11} it was shown that 
the mean first passage time (MFPT) to the origin for a single
diffusive searcher
becomes finite in the presence of resetting 
(in contrast  to a purely diffusive search where the
MFPT diverges).  Moreover the MFPT has a
minimum value as a function of the resetting rate $r$ to the
fixed initial  position $x_0$. Thus, there is
an  optimal resetting  rate  $r$ as a function of the distance to the target
$x_0$.

In this work we address  the  question of  resetting strategies
which optimise the MFPT in a wider context. 
To this end, we make several  generalisations
of single-particle diffusion with resetting studied in 
\cite{EM11}. First,  we consider
a space dependent resetting rate $r(x)$.
Second, we consider
resetting to a {\em random} position 
$z$ (rather than a fixed $x_0$) drawn from  a resetting distribution 
${\cal P}(z)$. Finally, we consider
a probability distribution
for the absorbing target  $P_T(x)$.
The general question we ask is: what are the optimal
functions $r(x)$, ${\cal P}(x)$
that minimise the MFPT for a given
$P_T(x)$?
Although we do not propose a general solution, the examples
we study turn up some surprising results
and illustrate that answers to the problem may be non trivial.

The paper is organised as follows. In section 2 we review the calculation of the mean first
passage time for one-dimensional duffusion in the presence
of resetting  to the intial position with rate $r$. In section 3 we introduce
spatial dependent resetting $r(x)$ and work out the example
of a non resetting window of width $a$ around the intial point.
In section 4 we consider
the  generalisation to a resetting distribution ${\cal P}(z)$  
and to a distribution of the target site $P_T(x)$.
In section 5 we formulate  the general problem of optimising the 
mean first passage time with respect to the resetting distribution
${\cal P}(z)$. We consider the  example of an exponential
target distribution and show that there is a transition in the
optimal resetting distribution. We conclude in section 6.

\section{First passage time for
single particle diffusion with resetting}

\begin{figure}
\centerline{\includegraphics[width=10cm]{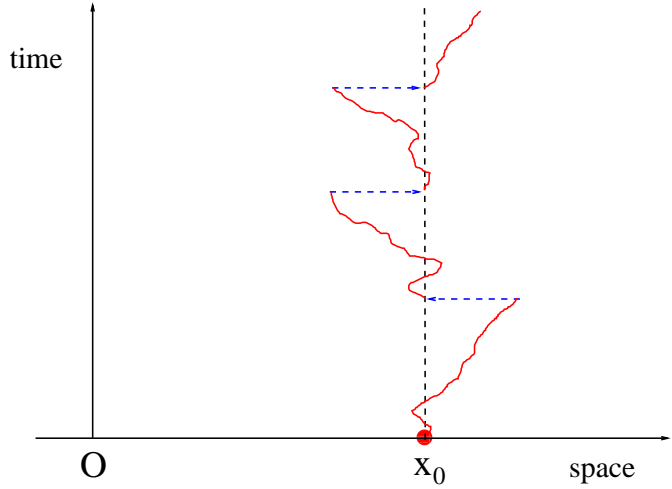}}
\caption{Schematic space-time trajectory of a one dimensional Brownian motion
that starts at $x_0$ and resets stochastically to its initial position
$x_0$ at rate $r$.}
\label{reset1.fig}
\end{figure}

We begin by briefly reviewing the one-dimensional 
case of diffusion with resetting 
to the initial position $x_0$ (see Fig. \ref{reset1.fig}), introduced in
Ref. \cite{EM11}.
The Master equation for 
$p(x,t|x_0)$, the  probability distribution for  the particle
at time $t$ having started from initial position $x_0$,  reads
\begin{equation}
\frac{\partial p(x,t|x_0)}{\partial t}
= D\frac{\partial^2 p(x,t |x_0)}{\partial x^2}
 - r p(x,t|x_0) + r\delta(x-x_0)
\label{me}
\end{equation}
with initial condition $p(x,0|x_0) = \delta(x-x_0)$.  In Eq. (\ref{me}) $D$
is the diffusion constant of the particle and $r$ is the resetting
rate to the initial position $x_0$. The second term on the right hand side (rhs)
of Eq. \ref{me} denotes the loss of probability from the position $x$ due
to reset to the initial position $x_0$, while the third term denotes
the gain of probability at $x_0$ due to resetting from all other positions. 

The stationary state of (\ref{me}) is the solution of
\begin{equation}
D\frac{\partial^2 p^*(x |x_0)}{\partial x^2}
 - r p^*(x|x_0) =  -r\delta(x-x_0)
\label{mess}
\end{equation}
which is determined by the elementary Green function technique,
which we now recall.

The  solutions to the homogeneous counterpart of
(\ref{mess}) are ${\rm e}^{\pm \alpha_0 x}$
where   
\begin{equation}
\alpha_0 = \sqrt{r/D}\;.
\end{equation}
The solution to (\ref{mess}) is constructed from linear combinations
of these solutions which satisfy the following boundary conditions:
$p^* \to 0$ as $x \to \pm \infty$,
and $p^*$ is continuous at $x=x_0$.
Imposing these conditions yields  
\begin{equation}
p^*(x|x_0) = A\exp( - \alpha_0 |x-x_0|)\;.
\label{pss2}
\end{equation}

Note that (\ref{pss2})  has a
cusp at $x= x_0$.
The constant $A$ is fixed  by the discontinuity of the first derivative at
$x=x_0$ which is determined by integrating (\ref{mess})
over a small region about $x_0$
\begin{equation}
\left. \frac{ \partial p^*(x|x_0)}{\partial x} \right|_{x \to x_0^+}
-\left. \frac{ \partial p^*(x|x_0)}{\partial x} \right|_{x \to x_0^-}
= -\alpha_0^2\;.
\end{equation}
Carrying  this out yields $A= \alpha_0/2$ so that
\begin{equation}
p^*(x|x_0) = \frac{\alpha_0}{2} \exp( - \alpha_0 |x-x_0|)\;.
\label{pss}
\end{equation}
Alternatively, the constant $A$ 
in (\ref{pss2}) could be fixed by the normalisation of the probability distribution (\ref{pss2}).

Note that (\ref{pss}) is a non-equilibrium stationary state by which
it is meant that there is circulation of probability even in the
one-dimensional geometry.  At all points $x$ there is always a
diffusive flux of probability in the direction away from $x_0$ given
by $ - D \partial p/\partial x$, and a nonlocal resetting flux in the
opposite direction from all points $x \neq x_0$ to $x_0$.

\subsection{Mean first passage time}
We now consider the mean first passage time for  the diffusing particle
to reach the origin.
One can think of an absorbing target at the origin which instantaneously
absorbs the particle (see e.g. \cite{Redner}).

A standard approach to first-passage problems is to use
the backward Master equation  where one 
treats the {\em initial}  position as a variable (for a review see Ref. 
\cite{BFreview}). 
Let $Q(x,t)$ denote the survival probability of the particle
up to time $t$ (i.e. the probability that the particle has not visited the origin
up to time $t$) starting 
from the initial position $x$. 
The boundary and initial conditions are $Q(0,t)=0$, $Q(x,0)=1$
(see e.g. \cite{ST79} for more general 
reaction boundary conditions).

The backward Master equation (where the
variable $x$ is now the initial position)
reads for the survival probability $Q(x,t)$ 
\begin{equation}
\frac{\partial Q(x,t)}{\partial t}
= D\frac{\partial^2 Q(x,t)}{\partial x^2}
 - r Q(x,t) + rQ(x_0,t)\;.
\label{bme}
\end{equation}
Note that $Q(x,t)$
depends implicitly on the resetting 
position $x_0$ due to the third term on the right hand side of
(\ref{bme}).
The second and third terms on the rhs correspond 
to the resetting of the {\em initial} position 
from $x$ to $x_0$,
which implies
a loss of probability from $Q(x,t)$ and a gain of probability
to $Q(x_0,t)$.

Equation (\ref{bme}) may be derived as follows.
We consider the survival probability $Q(x, t+\Delta t)$
up to time $t+ \Delta t$, where $\Delta t$ is a small interval of time.
We divide the time interval $[0,t+\Delta t]$ into two intervals: $[0,\Delta t]$
and $[t, t+\Delta t]$. In the first interval $[0, \Delta t]$, there are
two possibilities: (i) with probability $r\Delta t$, the particle
may be reset to $x_0$ and then for the subsequent
interval $[\Delta t, t+\Delta t]$ this $x_0$ will be the new starting position
and (ii) with probability $(1-r \Delta t)$, no resetting takes place, but instead
the particle diffuses to a new position $(x+\xi)$ in time $\Delta t$ where $\xi$ 
is a random variable
distributed according to a gaussian distribution 
$P(\xi) = (4 \pi D \Delta t)^{-1/2} \exp (- \xi^2/4D \Delta t)$. This
new position $(x+\xi)$ then becomes the starting position for the subsequent
second interval $[\Delta t, t+\Delta t]$. One then sums over all possible values
of $\xi$ drawn from $P(\xi)$. Note that we are implicitly using the
Markov property of the process whereby for the second interval $[\Delta t, 
t+\Delta t]$, only the end position of the first interval $[0,\Delta t]$ matters.  
Taking into acount these two possibilities, one then gets
\begin{equation}
\fl Q(x, t +\Delta t)
= r\,\Delta t Q(x_0,t)
  +(1-   r\, \Delta t )\int \D \xi P(\xi) Q(x+\xi,t)
\end{equation}
which can be rewritten as
\begin{equation}
\fl \frac{Q(x, t +\Delta t) - Q(x,t)}{\Delta t}
= \int \frac{\D \xi}{\Delta t} P(\xi)( Q(x+\xi,t)-Q(x,t)) 
  + r Q(x_0,t) - r Q(x,t) + O(\Delta t)\;.
\end{equation}
Taking the limit $\Delta t \to 0$ 
then yields (\ref{bme}).

The mean first passage time  $T$
to the origin  beginning from position $x$
is obtained by noting that
$-\frac{\partial Q(x,t)}{\partial t} \D t$
is the probability of absorption by the target in time
$t \to t + \D t$.
Therefore, on integrating by parts, we have
\begin{eqnarray}
T = -\int_0^\infty t 
\frac{\partial Q(x,t)}{\partial t} \D t 
 =  \int_0^\infty Q(x,t) \D t
\end{eqnarray}
(assuming that $t Q(x,t) \to0$ as $t \to \infty$). Integrating
(\ref{bme}) with respect to time  yields
\begin{equation}
-1  =  D\frac{\partial^2 T(x)}{\partial x^2}
 - r T(x)  + rT(x_0)
\label{Tme}
\end{equation}
with boundary conditions $T(0)=0$
and $T(x)$ finite as $x \to \infty$.

To solve for the mean first passage time beginning at the resetting position 
$x=x_0$  we first consider the initial
position to be at $x>0$, different from the resetting position $x_0$,
then solve (\ref{Tme}) with arbitrary $x$ and $x_0$.
Once we have this solution  we   set
$x=x_0$ to determine $T(x_0)$
self-consistently.

The general solution to (\ref{Tme}) is
\begin{equation}
T(x) 
 = A {\rm e}^{\alpha_0 x} + B {\rm e}^{-\alpha_0 x} 
+\frac{ 1 + rT(x_0)}{r}
\end{equation}
where $\alpha_0= \sqrt{r/D}$.
The boundary condition that  $T(x)$ is finite as $x\to \infty$ implies
$A=0$ and the boundary condition $T(0) =0$ fixes $B$.
Thus
\begin{equation}
T(x) 
 = 
\frac{ 1 + rT(x_0)}{r}\left[
1-{\rm e}^{-\alpha_0 x} \right]
\;.
\end{equation}
Solving for $T(x_0)$ self-consistently yields 
\begin{equation}
T(x_0) = \frac{1}{r} \left[\exp(\alpha_0 x_0) -1\right]= \frac{1}{r} 
\left[\exp\left(\sqrt{r/D}\, x_0\right)-1\right]\;.
\label{Tres}
\end{equation}

Note from (\ref{Tres}) that, for fixed $x_0$, $T$ is finite for $0<r<\infty$.
As a function of $r$ for fixed $x_0$, $T$ diverges
when $r \to 0$ as 
\begin{equation}
T \simeq   \frac{x_0}{ (D r)^{1/2}} \;.
\end{equation}
This is expected since as $r\to 0$, one should recover the pure diffusive
behaviour (no resetting) for which the $T$ is divergent--due to the
large excursions that take the diffusing particle away from the
target at the origin.
Also $T$ diverges rapidly as $r \to \infty$, the explanation
being that as the reset rate increases the diffusing particle 
has less time between resets to reach the origin. In other words,
the high resetting rate to $x_0$ cuts off the trajectories that
bring the diffusing particle towards the target. 
\begin{figure}
\includegraphics[width=.7\hsize]{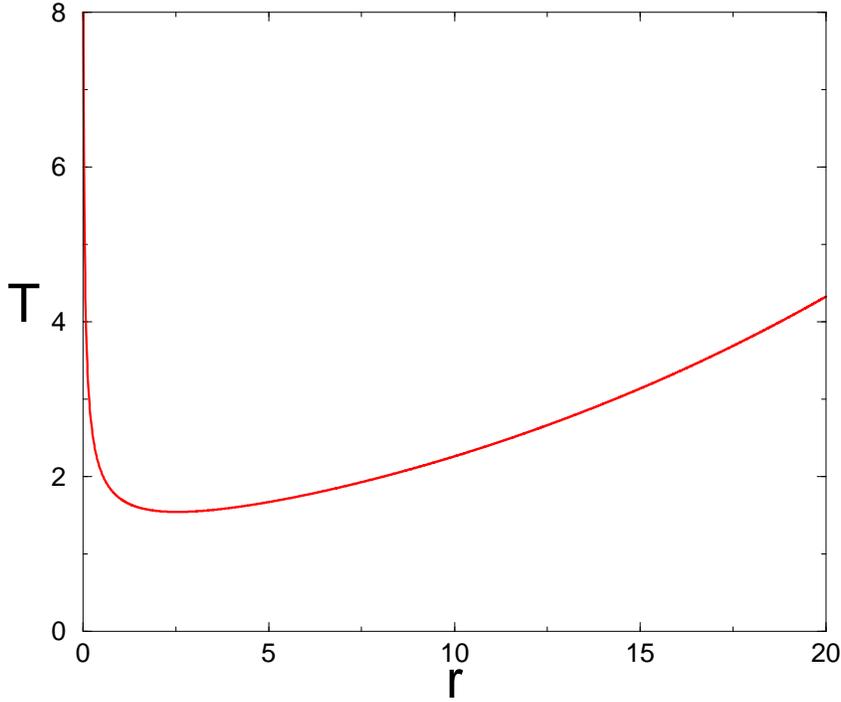}
\caption{The mean first passage time 
$T=\frac{1}{r}\left[\exp\left(\sqrt{r/D}\,x_0\right)-1\right]$ plotted as a 
function of $r$ for fixed $x_0=1$ and $D=1$. clearly $T$ diverges
as $r\to 0$ and as $r\to \infty$ with a single minimum at $r^*=2.53962\ldots$.}
\label{mfpt.fig}
\end{figure}

We now consider $T$ as a function of $r$ for a given value of $x_0$
and define the reduced variable 
\begin{equation}
z = \alpha_0 x_0 = \left(\frac{r}{D}\right)^{1/2} x_0.
\label{zdef}
\end{equation}
Since  $T$ diverges as $r \to 0$ and $r\to \infty$
it is clear that there must be a minimum of $T$ with
respect to $r$ (see Fig. \ref{mfpt.fig}). 
The condition for the minimum, $\displaystyle \frac{\D T}{\D r}=0$, reduces  to
the transcendental equation 
\begin{equation}
\frac{z}{2} = 1 - {\rm e}^{-z}
\label{zopt}
\end{equation}
which has a unique non-zero
solution $z^*= 1.59362...$. In terms of the restting rate, this means 
an optimal resetting rate $r^*=(z^*)^2 
D/x_0^2= (2.53962 \ldots) 
D/x_0^2$, for which the mean first passage time $T(x_0)$ is minimum.
The dimensionless variable $z$ (\ref{zdef})
is a ratio of two lengths: $x_0$, the distance 
from the resetting point to the target, and
$(D/r)^{1/2}$, which is the typical distance diffused between resetting events.
Thus, for fixed $D$ and $x_0$ the mean first passage time of the particle can
be minimised by choosing $r$ so that this ratio
takes the value $z^*$.

\section{Space-dependent resetting rate}

In this section we generalise the model of section 2
to the case of space-dependent  resetting rate $r(x)$.

The  master equation for the  probability distribution
$p(x,t|x_0)$  is generalised from (\ref{me}) to
\begin{equation}
\frac{\partial p(x,t|x_0)}{\partial t}
= D\frac{\partial^2 p(x,t |x_0)}{\partial x^2}
 - r(x) p(x,t|x_0) +\int \D x' r(x') p(x',t|x_0) \,\delta(x-x_0)
\label{rme}
\end{equation}
The third term on the right hand side now
represents the flux of probability
injected at $x_0$ through resetting from all points
$x \neq x_0$.

The stationary distribution $p^*(x|x_0)$ satisfies
\begin{equation}
D\frac{\partial^2 p^*(x |x_0)}{\partial x^2}
 - r(x) p^*(x|x_0) =  -\int \D x' r(x') p^*(x'|x_0) \,\delta(x-x_0)\;.
\label{rss}
\end{equation}

In general  the stationary state is difficult to determine
unless $r(x)$ has some simple form.
The equation for the mean first passage time becomes
\begin{equation}
-1  =  D\frac{\partial^2 T(x)}{\partial x^2}
 - r(x) T(x)  + r(x) T(x_0)
\label{Tmegen}
\end{equation}
Again, this is difficult to solve generally for arbitrary $r(x)$.

In the following we consider  a solvable example
where $r(x)$ is zero in a window around $x_0$ and is constant outside this
window.

\subsection{Example of a non-resetting window}
We  consider the case of a non-resetting
window of width $a$ about $|x_0|$, within which the  resetting process does not occur:
\begin{eqnarray}
r(x) &=& 0 \quad\mbox{for}\quad |x-x_0| < a \\
 &=& r \quad\mbox{for}\quad |x-x_0| \geq a\;.
\end{eqnarray}
This choice is a rather natural one in the sense that a typical searcher usually
doesn't reset when it is close to its starting point, but rather the
resetting event occurs when it diffuses a certain threshold distance
$a$ away from its initial position. 

The Master equation reads
\begin{eqnarray}
\frac{\partial p(x,t|x_0)}{\partial t}
&=& D\frac{\partial^2 p(x,t|x_0)}{\partial x^2}
  + r h(t) \delta(x-x_0)\qquad |x-x_0| <a \label{merx1}\\
&=& D\frac{\partial^2 p(x,t|x_0)}{\partial x^2}
  -rp(x,t|x_0)\qquad |x-x_0| \geq a \label{merx2}
\end{eqnarray}
where
\begin{equation}
h(t) =\int \D x\, p(x,t|x_0)\theta( |x-x_0|-a)\;,
\end{equation}
with initial condition $p(x,0) = \delta(x-x_0)$.
Thus $h(t)$ is  the probability that the particle
is outside the non-resetting window, i.e., in the resetting zone
at time $t$;
the particle is  reset to the origin with 
a total rate $h(t) r$.

First, we consider the 
stationary state.
One can solve for the
stationary probability using the Green function technique
of section 1.
For $|x-x_0| > a$ (outside the window), $p^*(x|x_0)$ satisfies
$ D\frac{\partial^2 p^*(x|x_0)}{\partial x^2}
   = rp^*(x|x_0)$ and should tend to zero
as $|x| \to \infty$.
For $0< |x-x_0| < a$ (inside the window), $p^*(x|x_0)$ satisfies
$ D\frac{\partial^2 p^*(x|x_0)}{\partial x^2}
   = 0$ for all $x\ne x_0$. The solution should be continuous at $x=x_0$, but
its derivative must undergo a jump at $x=x_0$ and the jump
discontinuity can be computed by integrating Eq. \ref{merx1} across
$x=x_0$.

Thus, noting that the solution should be symmetric about
$x=x_0$, one has
\begin{eqnarray}
p^*(x|x_0) &=& A \exp -\alpha_0( |x-x_0| -a)\qquad \mbox{for}\quad |x-x_0| > a 
\label{solrx1}\\
&=&    A - B\left( |x-x_0|-a\right))\qquad
\mbox{for}\quad |x-x_0| < a
\label{solrx2}
\end{eqnarray}
where $\alpha_0=\sqrt{r/D}$ and the constants $A$ and $B$ are determined by
the discontinuity in the derivative of
$p^*(x|x_0)$ at $x=x_0$ and  
the continuity of the derivative at $|x-x_0| = a$.

\begin{figure}
\includegraphics[width=.7\hsize]{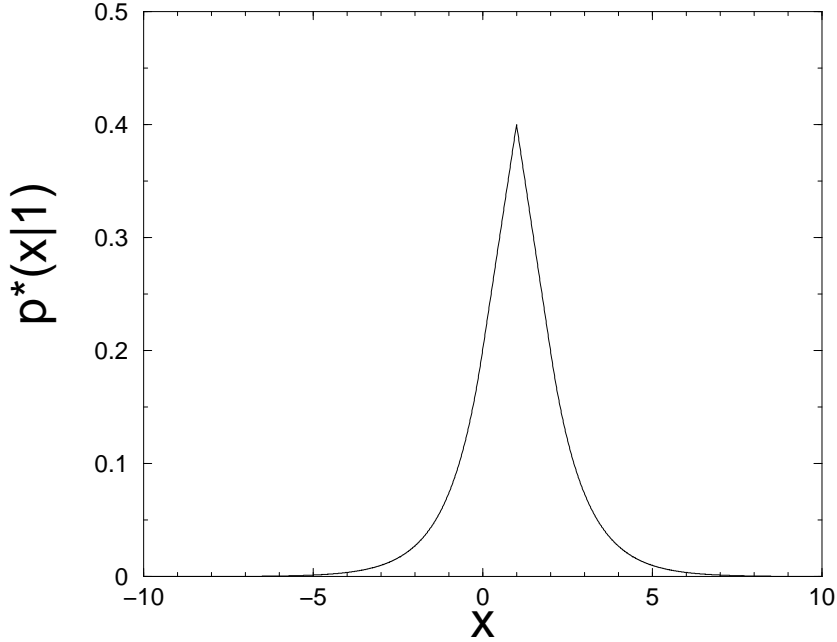}
\caption{The stationary solution $p^*(x|x_0)$ in Eq. \ref{solrx1}-\ref{solrx2} 
plotted as a 
function of 
$x$,
for the choice $x_0=1$, $a=1$, $r=1$ and $D=1$. The nonresetting window
is over $x\in [0,2]$ with the initial position at $x_0=1$. The solution
is symmetric around $x_0=1$ with a cusp at $x=x_0=1$.} 
\label{box.fig}
\end{figure}

The result is
\begin{equation}
A =\frac{\alpha_0^2}{2+2a\alpha_0 + a^2 \alpha_0^2}\qquad B= \alpha_0 A
\end{equation}
The solution has a cusp at $x= x_0$ and a discontinuity
in the second derivative at
$|x-x_0| = a$ (see Fig. \ref{box.fig}).

We now consider an absorbing trap at the origin.
The backward equation for $T(x)$, the mean time to absorption
beginning from $x$, reads
\begin{eqnarray}
\label{Twin1}
-1  =  D\frac{\partial^2 T(x)}{\partial x^2}
 - r T(x)  + rT(x_0) \qquad \mbox{for}\quad |x-x_0| > a\\
\label{Twin2}
-1  =  D\frac{\partial^2 T(x)}{\partial x^2}
 \qquad \mbox{for}\quad |x-x_0| < a\;.
\end{eqnarray}
The general solution to
(\ref{Twin1},\ref{Twin2}) is
\begin{equation}
T(x) = A x + B -\frac{x^2}{2D}
\end{equation}
and the  solution  that does not diverge as $x \to \infty$ is
\begin{eqnarray}
T(x) = \frac{1+rT(x_0)}{r} + C {\rm e}^{-\alpha_0(x-x_0-a)}
 \qquad \mbox{for}\quad x>x_0 +a \\
T(x) = \frac{1+rT(x_0)}{r} + E
{\rm e}^{-\alpha_0 x }
+ F
{\rm e}^{\alpha_0 x}
 \qquad \mbox{for}\quad x<x_0 - a \;.
\end{eqnarray}
The constants $A,B,C,E,F$ are determined by the continuity of $T(x)$ and
$T'(x)$ at $|x-x_0|=a$
and the boundary condition $T(0)=0$.
The result for $T(x_0)$ is
\begin{eqnarray}
T(x_0)& = \frac{1}{r(1+a\alpha_0)}&
\left[ \cosh \alpha_0(x_0-a) \left( 1 + 2 a \alpha_0 + \frac{3 a^2 \alpha_0^2}{2} + a^3 \alpha_0^3 \right)\right. \nonumber\\
&& +\left.\sinh \alpha_0(x_0-a) \left( 1 + 2 a \alpha_0 + \frac{3 a^2 \alpha_0^2}{2}
\right) \right] -\frac{1}{r}\;.
\end{eqnarray}
We now consider
the reduced parameters
$z = \alpha_0 x_0$ and $ y = \alpha_0 a$,
and $T$ as a function $y$  for $z$ fixed.
The allowed values of $y$ are $ 0 \leq y\leq z$.
At $y =  z$, one can show that 
$ \left. \frac{\D  T}{\D y}\right|_{y=z} > 0$.
Therefore the minimum of $T$ with respect to $y$
is either at $y= 0$ or at 
a non trivial minimum $ 0 \leq y\leq z$.

The condition for a minimum
$ \frac{\D  T}{\D y}= 0$ reduces to
\begin{equation}
\frac{2+y}{4 + 5y +2y^2}
=\tanh(z-y)
\end{equation}
Therefore the condition for there to be a nontrivial minimum
for $y>0$ is given by   $\tanh  z  > 1/2$
or equivalently  $z > (\log 3)/2 = 0.5493\ldots$.

In summary, 
the analysis of the condition for $T(y)$ to be
a minimum reveals that:
if $z < (\log 3)/2$ then $y=0$ is the minimum of $T(y)$;
if $z > (\log 3)/2$ then $T(y)$ has a nontrivial minimum at
$ 0 < y <z$.
Therefore, when $z < (\log 3)/2$ the introduction of
a window around the initial site
where resetting does not take place 
 does not reduce the mean time
to absorption.
A strategy of introducing 
a non-resetting window is  an effective one only when
the initial point is sufficiently far from the search
target. Otherwise it is advantageous to always reset.

\subsection{Optimal resetting function}
Having seen in the previous example that non-trivial behaviour emerges
for a simple spatial-dependent resetting rate $r(x)$, one can ask for
the optimal function $r(x)$.  The optimisation problem would be to minimise
$T$ under certain constraints pertaining to the information available
to the searcher. Clearly if there are no constraints, that is one can
use full information about the target position, the optimal strategy
is to reset immediately whenever $x>x_0$ and not reset when $x<x_0$.
This corresponds to the choice
\begin{eqnarray*}
r(x) &=& 0 \quad \mbox{for}\quad x <x_0 \\
r(x) &=& \infty \quad \mbox{for}\quad x >x_0\;.
\label{ropt}
\end{eqnarray*}
In this case  problem (\ref{Tmegen}) reduces to the mean first passage 
time of a diffusive particle with reflecting barrier at $x_0$ 
the solution of which is
\begin{equation}
T^*(x_0) =  \frac{x_0^2}{2D}\;.
\label{opt}
\end{equation}
Thus, (\ref{opt}) gives the lowest possible mean first passage time
for a diffusive process.  One can then ask about how close simple
strategies, such as a spatially constant resetting rate $r$ or
non-resetting window, come to approaching this bound.

For example, the case of spatially constant resetting rate $r$ considered
in section 2 yields a minimum MFPT using (\ref{zopt})
\begin{equation}
T= \frac{x_0^2}{D} \frac{({\rm e}^z-1)}{z^2} =
\frac{x_0^2}{2D} \frac{{\rm e}^{z^*}}{z^*} = 3.0883...\, T^*(x_0)
\end{equation}
As noted in section 3.1 the value  3.0883 may be improved upon by considering
a non-resetting window around $x_0$.

However, (\ref{opt}) uses the crucial information of whether the target
(at $x=0$) is to the right or left of the resetting site $x_0$.  More
realistically, the searcher would not have this information. 
The relevant optimisation problem is  to
find the optimal resetting rate $r(|x-x_0|)$ (constrained to be a function of the distance $|x-x_0|$ from the resetting site) that minimises
$T(x_0)$. This remains an open problem.

\section{Resetting distribution and target distribution}

In this section we consider the generalisation to a system with
resetting to points distributed according to ${\cal P}(z)$. We shall
also consider a distribution of the target site $P_T(x)$.

\subsection{Stationary state}

We begin by considering again the one-dimensional 
case of diffusion but this time with resetting 
to a random position:
at rate $r$ the particle is reset to a random position $z\to z + \D z$
drawn with probability
${\cal P}(z) \D z$. We refer to
${\cal P}(z)$ as the reset distribution.
For simplicity we take the initial position $x_0$ to be distributed
according to the same distribution as the reset position $p(x_0,0) =
{\cal P}(x_0)$.

The  Master equation
for the probability density $p(x,t)$ now reads
\begin{equation}
\frac{\partial p(x,t)}{\partial t}
= D\frac{\partial^2 p(x,t) }{\partial x^2}
 - r p(x,t) + r {\cal P}(x)\;.
\label{meP}
\end{equation}

The stationary solution to
(\ref{meP}) is simply found using
(\ref{pss}) as the Green function:
\begin{equation}
p^*(x) = \int \D z \,
{\cal P}(z) p^*(x|z)
\label{p*}
\end{equation}
which, using  $p^*(x|x_0)$  given by
(\ref{pss}),
 yields
\begin{equation}
p^*(x) = \frac{\alpha_0}{2}\int \D z \,
{\cal P}(z) \exp(-\alpha_0|x-z|))\;.
\label{p*2}
\end{equation}

\subsection{Mean first passage time}
The mean first passage time, $T(x_0,x_T)$,  to a target point $x_T$,
starting from $x_0$ with resetting distribution
${\cal P}(z)$,
satisfies
\begin{equation}
-1  =  D\frac{\partial^2 T(x_0,x_T)}{\partial x_0^2}
 - r T(x_0,x_T)  + r\int \D z {\cal P}(z)\,T(z,x_T)
\label{TmePz}
\end{equation}
with boundary condition $T(x_T,x_T)=0$.
To solve this equation we let
\begin{equation}
F(x_T)=\int \D z\, {\cal P}(z)T(z,x_T)
\end{equation}
then write down the general solution to
(\ref{TmePz}) and solve 
for $F(x_T)$ self-consistently.

The general solution of
(\ref{TmePz}) which is 
finite as $x_0 \to \infty$ is
\begin{equation}
T(x_0,x_T) 
 = A {\rm e}^{ -\alpha_0| x_0-x_T|} +
\frac{1}{r}+F(x_T)
\label{Tgen}
\end{equation}
The boundary condition $T(x_T,x_T)=0$ implies
$A = - \left( \frac{1}{r} + F \right)$.
Then  substituting this expression for $A$ in 
(\ref{Tgen}) and integrating we find
\begin{equation}
F(x_T) = \left(
\frac{1}{r}+F(x_T)\right)
\left(1-  \int \D z {\cal P}(z)e^{-\alpha_0|z-x_T|} \right)
\end{equation}
which yields
\begin{equation}
F(x_T) = 
\frac{1}{r}
\left(\frac{\alpha_0}{2 p^*(x_T)}-1\right).
\end{equation}
Inserting this into (\ref{Tgen})  we obtain
\begin{equation}
T(x_0,x_T) = \frac{\alpha_0}{2 r p^*(x_T)} \left[1- \exp(-\alpha_0| x_0-x_T|) \right]\;.
\end{equation}
As noted above it is convenient to choose
the same distribution for $x_0$ as the resetting distribution.
Averaging over $x_0$ then gives using (\ref{p*})
\begin{equation}
T(x_T) = \frac{1}{ r }
\left[ \frac{\alpha_0}{2  p^*(x_T)} -1 \right]\;.
\label{TxT}
\end{equation}
Equation (\ref{TxT}) gives the expression for
the mean first passage time to a target positioned at $x_T$.
Let us check  the case of a single 
position $x_0$ to which the particle is reset
$ {\cal P}(z) = \delta(z-x_0)$.
In this case (\ref{TxT}) becomes
\begin{equation}
T(x_T) = \frac{1}{r}
\left[ \frac{\alpha_0}{2} \frac{1}{p^*(x_T)}  -1 \right]
\end{equation}
which recovers (\ref{Tres}) when $x_T$ is set to 0.

Finally, we   average  over possible target positions 
drawn from a target distribution:
$P_T(x_T)$
\begin{equation}
\overline{T} = \frac{1}{r}
\left[ \frac{\alpha_0}{2} \int \D x_T \frac{ P_T(x_T)} {p^*(x_T)}  -1 \right]\;.
\label{Tav}
\end{equation}
Equation (\ref{Tav}) gives the main result of this section---
the MFPT for a resetting distribution ${\cal P}(x_0)$
and averaged over target distribution $P_T(x_T)$.

\section{Extremisation of mean first passage time}
Let us now consider the problem of extremising $\overline{T}$
given by (\ref{Tav}), for a given target distribution $P_T(x)$, with
respect to the resetting distribution ${\cal P}(z)$.
Throughout this section we will assume a symmetric target distribution:
$P_T(x)=P_T(-x)$
and $P'_T(x)=-P_T'(-x)$.

The problem is to minimize
the functional appearing in (\ref{Tav}):
$\int \D x  \frac{ P_T(x)}{p^*(x)}$
where 
\begin{equation}
p^*(x) = \frac{\alpha_0}{2} \int \D z\, {\cal P}(z)
{\rm e}^{-\alpha_0|z- x|}\;,
\label{pssP}
\end{equation}
subject to the constraint
$\int \D z {\cal P}(z) =1$.
The functional derivative to be satisfied  is  
\begin{equation}
\frac{\delta}{\delta {\cal P}(y)}
\left[
 \int \D x  \frac{P_T(x)} {p^*(x)}
+ \lambda \int \D x\,  {\cal P}(x)\right] =0
\label{fd}
\end{equation}
where $\lambda$ is a Lagrange multiplier.
Condition (\ref{fd})  yields 
\begin{equation}
\int 
\D x \frac{P_T(x)}{\left[p^*(x)\right]^2}
{\rm e}^{-\alpha_0|y-x|} =\frac{2\lambda}{\alpha_0} \;.
\label{EL}
\end{equation}
For (\ref{EL}) to hold for all $y$
requires that
\begin{equation}
\frac{P_T(x)}{\left[p^*(x)\right]^2} = \lambda\;,
\label{pssopt1}
\end{equation}
or fixing $\lambda$ through the normalisation of
$p^*(x)$
\begin{equation}
p^*(x) = \frac{P^{1/2}_T(x)}{\int \D z P^{1/2}_T(z)}\;.
\label{pssopt}
\end{equation}

Equation (\ref{pssopt})   implies that
to minimise $T$
the  stationary probability distribution should be given by the
square root of the target distribution.
This result has been derived in \cite{Snider11}
for the case 
of searching for the target by sampling a probability distribution
${\cal P}(x)$. This corresponds to the limit $r\to \infty$ of our model.
For $r <\infty$ we have the additional constraint that
the optimal
$p^*$ should be realisable
from a resetting distribution ${\cal P}(z)$ through formula (\ref{pssP}).

Equation (\ref{pssP}) may be solved for
${\cal P}(z)$ for a desired  $p^*(x)$ 
by taking the Fourier transform and using the convolution theorem
to give
\begin{equation}
{\cal \widetilde{P}}(k) = \left(1+\frac{k^2}{\alpha_0^2} \right)
\widetilde{p}^*(k)
\label{Pkopt}
\end{equation}
where 
${\cal \widetilde{P}}(k)$ is the Fourier transform
of ${\cal {P}}(x)$ and
$\widetilde{p}^*(k)$  is the Fourier transform of
$p^*(x)$.

We  may invert the Fourier transformation to find
\begin{equation}
{\cal P}(x) = p^*(x) -\frac{1}{\alpha_0^2} \frac{\D^2 p^*(x)}{\D x^2}\;.
\label{Popt}
\end{equation}
However this solution may become negative
in which case the solution to the optimisation problem
is unphysical.

\subsection{Example of an exponential target distribution}
As a simple example, we  consider an exponentially decaying target distribution
peaked  at $x=0$:
\begin{equation}
P_T(x) = \frac{\beta}{2} {\e }^{-\beta|x|}\;.
\label{PTexp}
\end{equation}
We first note that for a delta function resetting distribution
${\cal P}(z) = \delta(z-x_0)$ the
mean first passage time (\ref{Tav}) diverges when 
$\alpha_0 >\beta$.
Therefore, for small $\beta$ (a broad target distribution)
one expects an optimal resetting distribution (for fixed $\alpha_0$)
that differs from a delta function.

For $\beta < 2 \alpha_0$,
the optimal 
stationary distribution is from (\ref{pssopt})
\begin{equation}
p^*(x) = \frac{\beta}{4} {\e }^{-\beta|x|/2}
\end{equation} 
This expression  yields from (\ref{Popt}) a resetting
distribution that is always positive,  thus 
the optimal resetting distribution
\begin{equation}
{\cal P}(z) = \frac{\beta}{4} {\e }^{-\beta|z|/2}\left[ 1- \frac{\beta^2}{4\alpha_0^2} 
\right] + \frac{\beta^2}{4 \alpha_0^2} \delta(z)
\label{Poptexp}
\end{equation}

For $\beta > 2 \alpha_0$,
(\ref{Poptexp}) always gives negative probabilities
due to the first term.
Therefore we anticipate that 
${\cal P}(x) = \delta(x)$
is at least a locally optimal solution.
In fact one can prove this is the case by showing that any
distribution of the form
${\cal P}(x) = (1-\epsilon)\delta(x) + \epsilon f(x)$,
where $f(x) \geq 0$  and $\int \D x f(x) =1$
leads to an increase in (\ref{Tav}) at first order in $\epsilon$
when $\beta > 2 \alpha_0$. (As the proof is straightforward
but  somewhat tedious
we did not include it here.)
Thus a  transition in the form
of the optimal resetting distribution,
from a single delta function to (\ref{Poptexp}),
occurs at $\beta = 2 \alpha_0$.

\subsection{Inversion of $p^*(x)$}
As noted above, the constraint  ${\cal P}(x) \geq 0$ 
means that 
the optimal $p^*(x)$ given by 
(\ref{pssopt}) may not be realisable
from a physical resetting distribution ${\cal P}(z)$.
We are therefore led to the  general question of when a desired stationary distribution  (e.g. (\ref{pssopt})) which we denote $g(x)$
may be generated from
(\ref{pssP})
i.e. when  can we invert
\begin{equation}
g(x) =
\frac{\alpha_0}{2} \int \D z {\cal P}(z)
{\rm e}^{-\alpha_0|z- x|}
\label{g(x)}
\end{equation}
to obtain a physical ${\cal P}(z)$?

Let us first discuss  a sufficient condition
for  the resetting distribution implied by (\ref{g(x)})
to be physical.

Equation (\ref{Pkopt}) relates the characteristic functions of the two
distributions ${\cal P}(x)$ and $g(x)$ (given there by $p^*(x)$).  
In terms of the
characteristic function,  Polya's theorem \cite{PT} 
states that if a function $\phi(k)$ satisfies: $\phi(0) = 1$; $\phi(k)$ is
even; $\phi(k)$ is convex for $k>0$, and $\phi(\infty) = 0$; 
then $\phi(k)$ is the
characteristic function of an absolutely continuous
symmetric distribution.
Polya's theorem therefore gives a sufficient
condition for ${\cal P}(x)$ implied by $p^*(x)$ to be physical.

The condition for convexity becomes in one dimension
\begin{equation}
\frac{\D^2}{\D k^2} \left[ \left(1 + \frac{k^2}{\alpha_0^2}\right) g(k)\right]
\geq 0\quad \mbox{for all}\quad k \geq 0\;.
\label{conv}
\end{equation}
If the function ${\cal \widetilde{P}}(k)$ does not satisfy the
conditions of Polya's theorem, 
the solution of (\ref{g(x)})  is invalid as a
probability distribution i.e. the desired $g(x)$ cannot be
realised from any resetting probability  distribution ${\cal P}(z)$.

In the case where
(\ref{g(x)})
may not be inverted to give a physical 
${\cal P}(x)$,
it may be possible to generate the desired form for $g(x)$ on a finite region
by choosing a compact support for ${\cal P}(z)$.
Let us assume $g(x)$ to be a symmetric function of $x$.
Then if we choose
\begin{eqnarray}
{\cal P}(z) &=& \lambda 
\left[ g(z) - \frac{1}{\alpha^2_0} \frac{\D^2 g(z)}{\D z^2}
\right]\quad\mbox{for}\quad |z| \leq y_0\\
&=& 0 \quad\mbox{for} \quad |z| > y_0\;,
\label{compsup}
\end{eqnarray}
where $\lambda$ is a normalising constant, we find
\begin{eqnarray}
p^*(x) &=& \lambda g(x) \quad\mbox{for}\quad |x| \leq y_0 \label{P*cs0}\\
p^*(x) &=& \lambda g(y_0){\rm e}^{\alpha_0(y_0 - |x|)}
 \quad\mbox{for}\quad |x| \geq y_0 
\label{P*cs}
\end{eqnarray}
provided that $y_0$ is chosen so that
\begin{eqnarray}
g(y_0) + \frac{1}{\alpha_0}g'(y_0) &=&0 \label{ydef0} \\
g(-y_0) - \frac{1}{\alpha_0}g'(-y_0) &=&0 \;.
\label{ydef}
\end{eqnarray}
(see \ref{app}).
The second condition follows from the first by the assumed symmetry of $g(x)$.
As an example, we consider the gaussian distribution
\begin{equation}
g(x) = \left( \frac{\beta}{\pi}\right)^{1/2}
{\rm e}^{-\beta x^2}
\end{equation}
The inversion of (\ref{g(x)}) using (\ref{Popt}) yields
\begin{equation}
{\cal P}(x)= g(x) -\frac{1}{\alpha_0^2} \frac{\D^2 g(x)}{\D x^2}
= \left( \frac{\beta}{\pi}\right)^{1/2}
{\rm e}^{-\beta x^2}\left[ 1 + \frac{2\beta}{\alpha_0^2} 
-\frac{4 \beta^2 x^2}{\alpha_0^2}\right]
\end{equation}
which becomes negative for
\begin{equation}
|x| >\frac{\alpha_0}{2\beta}\left(
1+ \frac{2\beta}{\alpha_0^2}\right)^{1/2}
\label{Pgauss}
\end{equation}
However, choosing a compact support for
${\cal P}(z)$ according to
(\ref{ydef}),
yields
\begin{equation}
y_0= \frac{\alpha_0}{2\beta}
\end{equation}
and we find that the resulting distribution
(\ref{Pgauss}) is positive for all $x$.

\section{Conclusion}
In this paper we have considered some generalisations of diffusion
with stochastic resetting to the case of spatial-dependent resetting rate and a
resetting distribution.  We have considered the mean first passage
time to a target which may be situated at a fixed point (the origin)
or distributed according to a distribution and derived the result
(\ref{Tav}). The minimisation of this quantity may then be formulated
as an optimisation problem of which we have studied some examples.

In particular we have seen some perhaps unexpected results.  First, the
introduction of a non-resetting window around a fixed resetting
position reduces the MFPT when the target is sufficiently far
away. This suggests that the optimal resetting distribution, in the case
where we consider a resetting rate that is symmetric
about the restting point, $r(|x-x_0|)$ may be non-trivial.  We have also seen
that in the case of an exponentially distributed target (\ref{PTexp})
the optimal resetting distribution undergoes a transition from
(\ref{Poptexp}) to a pure delta function at the origin.

Generally, the computation of an optimal resetting distribution is an
open problem since the resetting distribution that minimises
$\overline{T}$ may be become negative over some domain and therefore
nonphysical.  In the case where (\ref{Popt}) becomes unphysical,
although we do not have a solution to the extremisation problem of
minimising $\overline{T}$ subject to the additional constraint ${\cal
  P}(x)\geq 0$ we may propose likely candidates for extremal
solutions.  One possibility for the optimal physical solution is one
that has compact support i.e. since the constraint for the
distribution to be physical is that ${\cal P}(x)\ge 0$, one might
expect that the optimal solution lies on the boundary where ${\cal
  P}(x) = 0$ for some regions of $x$.  However, we have no proof that this is the
case.

Further considerations for optimising mean first passage times in a
more realistic search process would be to add a cost to resetting
since in the present model the diffusive particle instantaneously
resets to its selected resetting position.  This could be implemented
by attributing some time penalty to each resetting event, as is the
case in the framework intermittent searching.
\appendix
\section{Proof that (\ref{compsup}) yields (\ref{P*cs})}
\label{app}
We wish to show that expression (\ref{compsup})
for ${\cal P}(x)$
\begin{eqnarray}
{\cal P}(z) &=& \lambda 
\left[ g(z) - \frac{1}{\alpha^2_0} \frac{\D^2 g(z)}{\D z^2}
\right]\quad\mbox{for}\quad |z| \leq y_0 \label{comps}\\
&=& 0\quad\mbox{otherwise}\;,
\label{compsup2}
\end{eqnarray}
yields  (\ref{P*cs0})-(\ref{P*cs}) for the stationary distribution
given by (\ref{p*2}), provided that
(\ref{ydef0})-(\ref{ydef}) holds.

We begin by inserting (\ref{comps})-(\ref{compsup2}) into
(\ref{p*2}) in the case $|x|< y_0$:
\begin{eqnarray}
p^*(x)
&=& \frac{\alpha_0\lambda}{2}
\left\{
\int_{-y_0}^x\left[ g(z) -\frac{1}{\alpha_0^2}\frac{\D ^2 g(z)}{\D z^2}
\right]
{\rm e}^{-\alpha_0(x-z)} \right.\nonumber \\
&&+
\left. \int_{x}^{y_0}\left[ g(z) -\frac{1}{\alpha_0^2}\frac{\D ^2 g(z)}{\D z^2}
\right]
{\rm e}^{-\alpha_0(z-x)}
\right\}
\label{p*3}
\end{eqnarray}
We  use the following integration by parts,
valid for all $\alpha_0 \neq 0$
\begin{equation}
\fl \int_{a}^{b}\left[ g(z) -\frac{1}{\alpha_0^2}\frac{\D ^2 g(z)}{\D z^2}
\right] {\rm e}^{\alpha_0 z} \D z
= 
\left[ g(b) -\frac{1}{\alpha_0}\left.\frac{\D  g(z)}{\D z} \right|_{z=b}
\right]
\frac{{\rm e}^{\alpha_0b}}{\alpha_0}
-\left[ g(a) -\frac{1}{\alpha_0}\left.\frac{\D  g(z)}{\D z} \right|_{z=a}
\right]
\frac{{\rm e}^{\alpha_0a}}{\alpha_0}
\end{equation}
Inserting this into
(\ref{p*3}) and  cancelling terms yields
\begin{eqnarray}
p^*(x)
&=& \frac{\alpha_0 \lambda}{2}
\left\{ \frac{2g(x)}{\alpha_0}
+ \frac{{\rm e}^{-\alpha_0(x+y_0)}}{\alpha_0}
\left[ g(-y_0) -\frac{1}{\alpha_0}\left.\frac{\D  g(z)}{\D z} \right|_{z=-y_0}
\right] \right. \nonumber \\
&& + \left. \frac{{\rm e}^{\alpha_0(x-y_0)}}{\alpha_0}
\left[ -g(y_0) -\frac{1}{\alpha_0}\left.\frac{\D  g(z)}{\D z} \right|_{z=y_0}
\right]
\right\}
\end{eqnarray}
Then  conditions
(\ref{ydef0})-(\ref{ydef})  ensure that
$p^*(x) =  \lambda g(x)$ for $|x| < y_0$.

In the case $x > y_0$ we find
\begin{eqnarray}
p^*(x)
&=& \frac{\alpha_0\lambda}{2}
\int_{-y_0}^{y_0}\left[ g(z) -\frac{1}{\alpha_0^2}\frac{\D ^2 g(z)}{\D z^2}
\right]
{\rm e}^{-\alpha_0(x-z)}\nonumber \D z \\
&=&
 \frac{\alpha_0 \lambda}{2}
{\rm e}^{-\alpha_0x}
\left\{
\frac{{\rm e}^{\alpha_0y_0}}{\alpha_0}
\left[ g(y_0) -\frac{1}{\alpha_0}\left.\frac{\D  g(z)}{\D z} \right|_{z=y_0}
\right]
-
\frac{{\rm e}^{-\alpha_0y_0}}{\alpha_0}
\left[ g(-y_0) -\frac{1}{\alpha_0}\left.\frac{\D  g(z)}{\D z} \right|_{z=-y_0}
\right]
\right\}\nonumber \\
&=&
\lambda
{\rm e}^{-\alpha_0(x-y_0)}
 g(y_0)
\end{eqnarray}
where conditions
(\ref{ydef0})-(\ref{ydef}) have been used.

Similarly 
in the case $x < -y_0$ we obtain
$p^*(x) = \lambda g(-y_0){\rm e}^{\alpha_0(y_0 + x)}$.

\vspace{2em}
\section*{Acknowledgements}
MRE and SNM thank the Weizmann Institute for Weston Visiting Professorships

\end{document}